\documentclass[letterpaper, 10pt, conference]{ieeeconf}
\IEEEoverridecommandlockouts
\usepackage{cite}
\usepackage{amsmath,amssymb,amsfonts}
\usepackage[linesnumbered,vlined,algoruled]{algorithm2e}
\usepackage{graphicx}
\usepackage{float}
\usepackage{caption}
\usepackage{subcaption}
\usepackage{textcomp}
\usepackage{xcolor}
\def\BibTeX{{\rm B\kern-.05em{\sc i\kern-.025em b}\kern-.08em
\kern-.1667em\lower.7ex\hbox{E}\kern-.125emX}}
\usepackage{lipsum}

\usepackage{cleveref}
\usepackage{siunitx}
\usepackage{pifont}

\usepackage{tikz}
\usetikzlibrary{math}
\usetikzlibrary{calc}

\SetKwInOut{Parameter}{Parameter}

\newcommand{\key}[1]{\textbf{#1}}
\newcommand{\abr}{CC-PP}

\newcommand{\agent}{a}

    \title{Heuristic Planner for Communication-Constrained Multi-Agent Multi-Goal Path Planning}

    \author{
        J\'achym Herynek \\
        \textit{Czech Technical University in Prague} \\
        \textit{herynjac@fel.cvut.cz}
        \and
        Stefan Edelkamp \\
        \textit{Czech Technical University in Prague} \\
        \textit{Charles University} \\
        \textit{stefan.edelkamp@aic.fel.cvut.cz}
    }

\begin{document}

    \maketitle
    \thispagestyle{empty}
    \pagestyle{empty}

    \begin{abstract}

        In robotics, coordinating a group of robots is an essential task.
        This work presents the communication-constrained multi-agent multi-goal path planning problem and proposes a graph-search based algorithm to address this task.
        Given a fleet of robots, an environment represented by a weighted graph, and a sequence of goals, the aim is to visit all the goals without breaking the communication constraints between the agents, minimizing the completion time.
        The resulting paths produced by our approach show how the agents can coordinate their individual paths, not only with respect to the next goal but also with respect to all future goals, all the time keeping the communication within the fleet intact.

    \end{abstract}

    \begin{keywords}
        path-planning, mission planning, heuristic search, multi-agent path-finding
    \end{keywords}

    \section{Introduction}
        \label{sec_intro}

        As robotic systems are introduced in more areas of life and become more complicated, it is only natural that scenarios arise where large groups of agents need to coordinate.
        There are many ways the agents might interact with each other and their environment, and there are many limitations one might consider.
        This work is motivated by the constraint of limited communication distance.
        It establishes the problem of communication-constrained multi-agent multi-goal path planning (\abr) and offers a heuristic planner to solve the problem.

        The most straightforward problem considering coordination of multiple agents is the Multi-Agent Path-Finding~\cite{mapf} problem, which asks each agent to reach its goal without colliding with other agents.
        People walking on the street act in this manner: every participant has its own goal and only cares about not colliding with other actors.
        This is a well-known and well-studied problem but the fixed assignment of agents to goals can be somewhat limiting when considering a coordinated group of robots.

        Consider a scenario in which a group of robots needs to perform some simple action on several locations (such as collecting data).
        In a centralized system, a single arbiter plans the paths and actions for all the agents, and they execute them.
        The execution itself might be performed using very simple robots with lower computational resources.
        Without any additional constraints, this is the vehicle routing problem.
        However, if the robots can communicate with each other, one of them could play the role of the arbiter and transmit mission plans to the rest of the fleet.
        This way, one robot would be planning the paths and managing the gathered data for the whole fleet. As long as the communication remains unbroken, the whole system works as if each of the robots had access to the computational power of the arbiter.

        As a similar example, imagine a mother ship-style drone that sends out small drones.
        The smaller drones in such a scenario could act as its sensors and effectors and be almost bare-bones, and the system as a whole could have much larger potential and applicability by employing them.
        However, maintaining communication with the mother ship is critical; otherwise, the smaller drones might get lost.
        The \abr~ problem studied in this work could represent such a scenario.

        In contrast to the Multi-Agent Path-Finding problem, our the \abr~ problem assumes the goals are given in a sequence.
        As such, the goals are also not prescribed to specific agents.
        Since each agent can communicate its findings to the others, and the agents are interchangeable with respect to the goals, it does not matter which agent visits which goal.

        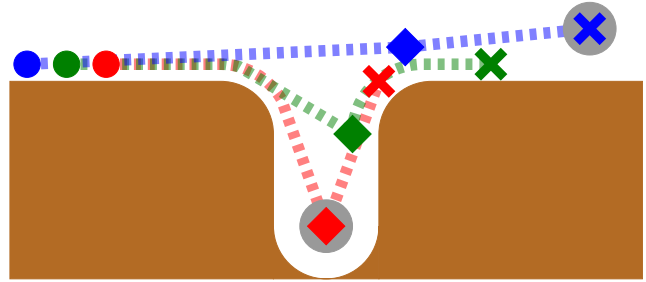
\begin{figure}
            \scalebox{1}[1]{
                \resizebox{\linewidth}{!}{
                    \begin{tikzpicture} [
                        obstacle/.style = {fill = orange!80!white!70!black, draw = orange!80!white!70!black},
                        goal/.style = {fill = black!40!white, draw = black!40!white},
                        agA/.style = {fill = blue, draw = blue},
                        agB/.style = {fill = green!50!black, draw = green!50!black},
                        agC/.style = {fill = red, draw = red},
                        agtrA/.style = {draw = blue, line width = 5, dashed, opacity = 0.5},
                        agtrB/.style = {draw = green!50!black, line width = 5, dashed, opacity = 0.5},
                        agtrC/.style = {draw = red, line width = 5, dashed, opacity = 0.5},
                    ]
                        \tikzmath{ \obsw = 4; \obsh = 3; \obsr = 0.8; \gap = 2*\obsr; }

                    \def \glr{0.4}
                    \def \agr{0.2}

                    \tikzmath{ \agdst = 0.2 + 2 * \agr; \agpos = 1.3 * \agr; }

                    \coordinate (gp1) at (\obsw + \obsr, \obsr);
                    \coordinate (gp2) at (2 * \obsw + \obsr, \obsh + 0.5 * \obsr + \glr);
                    \coordinate (agpA) at (\agpos, \obsh + \agpos);
                    \coordinate (agpB) at (\agpos + \agdst, \obsh + \agpos);
                    \coordinate (agpC) at (\agpos + 2 * \agdst, \obsh + \agpos);
                    \coordinate (agt1A) at (\obsw * 1.5, \obsh + 2 * \agpos);
                    \coordinate (agt1B) at (\obsw + 1.5 * \obsr, \obsh - \obsr);
                    \coordinate (agt1C) at (gp1);
                    \coordinate (agt2A) at (gp2);
                    \coordinate (agt2B) at (\obsw * 1 + 3.5 * \obsr + 1/2, \obsh + \agpos);
                    \coordinate (agt2C) at (\obsw + 2 * \obsr, \obsh);
                        \draw [obstacle]
                        (0, 0)  -- (\obsw, 0) -- (\obsw, \obsh - \obsr)
                        arc[start angle = 0, end angle = 90, radius = \obsr]
                        -- (0, \obsh) -- cycle;
                        \draw [obstacle]
                        (2 * \obsw + \gap, 0)  -- (\obsw + \gap, 0) -- (\obsw + \gap, \obsh - \obsr)
                        arc[start angle = 180, end angle = 90, radius = \obsr]
                        -- (2 * \obsw + \gap, \obsh) -- cycle;
                    \draw [obstacle]
                        (\obsw, \obsr) arc[start angle = -180, end angle = 0, radius = \obsr] --
                        (\obsw + \gap, 0) -- (\obsw, 0) -- cycle;
                    \draw[agtrC]
                        (agpC) -- (\obsw - \obsr, \obsh + \agpos)
                        arc[start angle = 90, end angle = 30, radius = \obsr + \agpos] -- (gp1);
                    \draw[agtrB]
                        (agpB) -- ((\obsw - \obsr, \obsh + \agpos)
                                arc[start angle = 90, end angle = 75, radius = \obsr + \agpos] --
                                (\obsw + 1.5 * \obsr, \obsh - \obsr);
                                \draw[agtrA]
                                (agpA) -- (\obsw * 1.5, \obsh + 2 * \agpos);

                                \draw[agtrC]
                                (agt1C) -- (\obsw + 2 * \obsr, \obsh);
                                \draw[agtrB]
                                (agt1B) arc [start angle = 180, end angle = 90, radius = \obsr + \agpos] --
                                (\obsw * 1 + 3.5 * \obsr + 1/2, \obsh + \agpos);
                                \draw[agtrA]
                                (agt1A) -- (gp2);

                                \draw[goal] (gp1) circle(\glr);
                                \draw[goal] (gp2) circle(\glr);

                                \draw[agA] (agpA) circle(\agr);
                                \draw[agB] (agpB) circle(\agr);
                                \draw[agC] (agpC) circle(\agr);

                                \draw[agA, rotate = 45] ($(agt1A) - (\agr, \agr)$) rectangle ($(agt1A) + (\agr, \agr)$);
                                \draw[agB, rotate = 45] ($(agt1B) - (\agr, \agr)$) rectangle ($(agt1B) + (\agr, \agr)$);
                                \draw[agC, rotate = 45] ($(agt1C) - (\agr, \agr)$) rectangle ($(agt1C) + (\agr, \agr)$);

                                \draw[agA, line width = 4pt] ($(agt2A) + (\agr, \agr)$) -- ($(agt2A) - (\agr, \agr)$);
                                \draw[agA, line width = 4pt] ($(agt2A) + (-\agr, \agr)$) -- ($(agt2A) - (-\agr, \agr)$);
                                \draw[agB, line width = 4pt] ($(agt2B) + (\agr, \agr)$) -- ($(agt2B) - (\agr, \agr)$);
                                \draw[agB, line width = 4pt] ($(agt2B) + (-\agr, \agr)$) -- ($(agt2B) - (-\agr, \agr)$);
                                \draw[agC, line width = 4pt] ($(agt2C) + (\agr, \agr)$) -- ($(agt2C) - (\agr, \agr)$);
                                \draw[agC, line width = 4pt] ($(agt2C) + (-\agr, \agr)$) -- ($(agt2C) - (-\agr, \agr)$);
                                \end{tikzpicture}
            } }
            \caption{Example behavior of the agents. \ding{108}: starting location, \ding{169}: locations at the time the first goal is reached, \ding{54}: final positions. While the red agent visits the first goal, the other two agents position themselves favorably with respect to the second goal, maintaining the communication distance as they do so.}
            \label{fig_illustration}
        \end{figure}

        Additionally, our work assumes a communication constraint on the fleet, which requires the agents to stay within communication distance of each other in a way that would make it possible to communicate with any agent in the fleet.
        Our work presents a formulation of the communication-constrained multi-agent path-finding problem and an algorithm for solving this problem, which is our main contribution.

        An illustration of the problem is shown in \Cref{fig_illustration}.
        It shows a situation where three agents need to visit two goals.
        While one of them makes the detour to the first goal, the other two are still able to improve their position with respect to the second goal without breaking the communication constraint.

        The paper is structured as follows:
        In \Cref{sec_related}, an overview of the current state of the art is presented.
        The studied \abr~problem is described in \Cref{sec_problem}, and the proposed approach is presented in \Cref{sec_method}.
        Our experiments and analysis of the solver are presented in \Cref{sec_results}.

    \section{Related Work}
        \label{sec_related}

        This section discusses the current state of the art with respect to \abr.
        As already mentioned, the two closest problems are Vehicle Routing~\cite{vrp} and Multi-Agent Path Finding~\cite{mapf}.

        First, let us take a look at the Traveling Salesman Problem (TSP)~\cite{tsp} and its generalization, the Vehicle Routing Problem~\cite{vrp}.
        The TSP consists of finding the shortest path in a complete graph that visits every node exactly once.
        This problem is very well studied, although it is NP-complete.
        Contemporary solvers, such as Concorde~\cite{tsp_concorde}, can optimally solve large instances (up to 85 000 nodes).
        Additionally, there are several polynomial approximation methods for certain variants of this problem, such as the 1.5-approximating Christofides algorithm~\cite{tsp_christofides}, which can be used for Euclidean TSP.

        The Vehicle Routing Problem, first introduced in~\cite{vrp}, originally called the Truck Dispatching Problem, is its variant in which several vehicles are considered (usually with a common depot to which all have to return).
        Since its formulation, many different variants of the problem have been proposed, such as capacitated vehicle routing, which imposes a restriction on the load of the vehicle~\cite{vrp_capacitated}, pickup and delivery, where items must be first picked up, before the customer is visited~\cite{vrp_pad}.
        As the name suggests, the vehicle routing problem is motivated by planning and dispatching deliveries using a fleet of vehicles.

        Multi-Agent Path Finding (MAPF)~\cite{mapf}, on the other hand, looks more closely at the paths the individual agents need to take.
        Instead of constructing high-level plans, the task here is to find a path for each agent, usually in a grid world, such that no two agents ever collide and each agent arrives at its target location.
        Usually, the objective to be minimized is either the sum of the path costs, or the makespan.
        The MAPF problem is NP-complete~\cite{mapf_np} as well.

        A naive approach to this problem is to consider the agents as a single unit and plan the path in the composite space.
        Each state in the composite state space consists of positions of all the individual agents, and the neighboring states are those states, obtained by any combination of moves of the agents.
        This approach scales poorly, since it is exponential in the number of agents.

        One commonly employed optimal approach is Constraint-Based Search~\cite{mapf_cbs}.
        This algorithm attempts to solve the problem by finding the shortest paths for all of the agents.
        Most likely, this will result in some number of conflicts (collisions).
        If some conflicts arise, the algorithm adds a constraint preventing that conflict and re-plans the shortest paths for the affected agents.
        Since there are multiple potential constraints that could prevent each collision, the algorithm branches.
        Once no more conflicts arise, the algorithm finishes all open branches that could still improve the solution and terminates.

        In addition to all the multi-agent path-finding research, communication-constrained planning has also seen some attention in the literature.
        However, most of the works concerning communication limits are focused on the exploration of unknown environments and on distributing the workload between the agents, which is quite different from the problem addressed in this paper.

        The authors of~\cite{los_ppcc} consider the communication constraint of keeping a line-of-sight chain between the agents.
        The line-of-sight constraint is kept with respect to the starting location.
        Multiple variants of the problem with different levels of a priori knowledge are considered.
        In the case where the agents have full knowledge of the environment, the problem boils down to arranging the agents along the path to the target.
        A similar problem is addressed in~\cite{los_a_ppcc}, where the agents require line-of-sight in order to localize themselves in the (a priori unknown) environment.

        In~\cite{explorer_relay_ppcc}, the authors propose a system in which agents assume one of two roles: explorers, which gather information, and relays, which support explorers and relay their information to the base.
        The agents form a communication tree, which can dynamically change whenever it is beneficial for the agents to switch roles.
        Additionally, the explorers may break the communication, explore outside of the range, and then bring the information back to a predetermined rendezvous point.
        The approach is further developed in~\cite{dynamic_role_ppcc}.
        A similar approach is also utilized in a decentralized manner in~\cite{relay_based_decentralised_ppcc}.

        A multi-robot exploration task with bounded distance between agents is addressed in~\cite{random_walk_ppcc}.
        The algorithm explores a randomly selected subset of the possible moves, thus reducing the dimensionality of the planning in the composite space of the agents.
        The best move from such a subset is selected based on a heuristic, which drives the agents towards unexplored frontiers.
        The algorithm also detects and resolves deadlocks, which can happen when the agents are driven in opposite directions and reach the limit of the communication distance.

        \cite{information_driven_ppcc} addresses the issue of data collection in unknown environments, in which the communication between the base station and between the individual agents is limited.
        Each agent selects its path based on communication availability and expected information gain in the selected location, and the agents attempt to coordinate together to avoid visiting the same location by multiple agents.
        In this work, communication of data is considered more as an objective rather than a constraint.

        Note that even though the TSP and Vehicle routing problems are conceptually close, in their formalization, they end up having very little in common with the studied communication-constrained path-planning.
        Still, the advances in this field allow us to use one of the TSP solvers to produce a reasonable ordering of the goals to give to our solver (this is considered to be given as a part of the scenario, as is discussed in \Cref{sec_problem}).

    \section{Problem Formulation}
        \label{sec_problem}

        The studied problem is that of finding the best possible plan for a group of agents to visit a sequence of goals in an obstacle-dense environment under the communication constraint.
        This part of the paper formally defines this problem.
        We assume a fleet of $n$ identical agents visiting $m$ distinct goals and communication limit $\lambda$.

        \subsection{Environment and Agent Movement}

            We assume an environment represented by a directed weighted graph $G = \left(V, E, w\right)$, where the weight of an edge represents the duration of the action associated with the movement from one vertex to the other.
            We also assume that there exists a distance function $\gamma: V \times V \rightarrow \mathbb{R}$ that computes the distance between any two vertices of the graph.
            The communication distance function $\gamma$ represents the communication distance between two agents (notice that $\gamma$ is only defined for vertices in the graph and not for the edges).

            Some of the vertices are assigned as \key{goals} (in a specified order): $g_{1}, \ldots, g_{m} \in V$ (where $m$ is the total number of goals).
            An agent \key{visits} goal $g_j$, if all goals $g_k, k < j$ have been visited and the agent is present at the associated vertex.
            Additionally, each agent has an associated starting vertex: $s_i \in V$.
            The starting location must not be in a collision and must not violate the communication constraint, otherwise the problem is infeasible.

            At any point in time, each agent occupies exactly one of the vertices, and no two agents may occupy the same vertex at the same time.
            Each agent that is not currently moving can move along an edge, which has a duration equal to the weight of that edge.
            This action takes effect instantaneously, meaning that the agent is considered to be on the target vertex for the whole duration of the action.
            This ensures that it is possible to evaluate $\gamma$ for any two agents at any time.

        \subsection{Communication constraint}

            The \key{communication constraint} is the key property of the studied problem.
            Informally, we assume that there is a limitation on the communication, which prevents agents from communicating if they are not within the communication limit, for example, if they are too far from each other.

            The constraint is defined using the distance function $\gamma$.
            The communication constraint imposes that any two agents must be able to communicate with each other, at least with other agents serving as repeaters.
            In our experiments, Euclidean distance was used.
            However, any arbitrary function would work as long as it is sufficiently fast to compute (as the computation of possible communication is the bottleneck of the algorithm).

            Assume agents $a$ and $b$ located at vertices $v_a$ and $v_b$, respectively.
            We define a relation $\mathcal{C}$, representing that the two agents are within the communication distance: $a~\mathcal{C}~b \iff \gamma\left(v_a, v_b\right) \leq \lambda$.
            We can then define the relation $\hat{\mathcal{C}}$ as the transitive closure of $\mathcal{C}$.
            This relation represents that two agents can communicate, either directly or with the help of other agents in the fleet.
            A group of $n > 2$ agents satisfies the communication constraint if any two agents can communicate, i.e., the relation $\hat{\mathcal{C}}$ holds for any two agents in the fleet.

        \subsection{Objective}

            The overall problem is then defined as follows:
            Given a weighted directed graph $G = \left(V, E, w\right)$, starting locations $\mathcal{S}$ of the agents, an ordered set of the goal vertices $\mathcal{G}$, distance function $\gamma$ and communication limit $\lambda$, the aim is to find a sequence of actions for each agent, such that the communication constraint is not violated, all of the goals are visited in the given order, and such that the completion time is minimized.

    \section{Proposed Method}
        \label{sec_method}
        
        The most straightforward solution to a problem like this is to use composite-state-space search.
        That means treating all the agents as a single complex robotic system and planning in the cross-product of their individual state spaces.
        However, the complexity of this method is exponential in the number of agents, which makes it computationally intractable even for small groups of agents.
        Our approach utilizes a best-first-search guided by a precomputed heuristic algorithm to compensate for the exponential complexity of this approach.

        The algorithm has two steps called \key{stages}.
        In the first stage, the heuristic is computed, and the second stage uses that heuristic to guide the greedy best-first-search.

        In both stages, the algorithm proceeds in discrete steps called \key{epochs}.
        Each epoch encompasses the computation between two goals, $g_i$ and $g_{i + 1}$.
        Since the paths between the goals are, to a large extent, independent of each other and the result of one epoch is the initial state of the next one, we can focus the search on the current path segment, and ignore everything else.

        \subsection{Stage 1}

            The first stage of the algorithm is described in \Cref{alg_CCS}, \Crefrange{alg_stage1_a}{alg_stage1_b}.
            In this stage, the algorithm restricts the search space and computes the heuristic for the second stage.
            Similarly to the whole computation, the heuristic can also be seen as a series of heuristics, each relating to a separate epoch.
            To compute the heuristic values for a given epoch, the algorithm selects one of the agents to reach the goal.
            For the other agents, it finds all the vertices they could be at when the goal is reached, and future epochs then consider those vertices as the potential starting locations.
            
            The computation of each epoch is described on \Crefrange{alg_foreach_epoch_s1}{alg_stage1_b}.
            During epoch $i$, one agent, called the \key{leader} ($\agent_{l_i}$), is selected to visit the goal $g_i$ (\Cref{alg_leader_select}).
            This is the only agent with a fixed target; that is, at the end of the epoch, it must hold that $\agent_{l_i}$ is at $g_i$.
            The other agents, \key{followers}, are free to wander around (position themselves favorably for the next epoch) as long as they keep the communication unbroken (note that the leader and follower relationship does not imply any hierarchy; if an agent is a leader, it only means that this agent is the one to visit the current goal).
            Since each epoch is considered on its own, the follower agents have no notion of the utility of their locations.
            Therefore, instead of determining a specific vertex to reach, the algorithm computes all the locations the agent could end at without breaking the communication.
            The result is then not a path but rather a set of possible locations at every time instance (called \key{result nodes}).

            During its search, each follower assumes that the other agents are located at the most favorable location possible.
            This means that even though each of the paths is achievable individually, not all combinations of paths (and indeed most of them) would be feasible together.
            This is why it is not possible to simply extract the paths from the first stage and produce a feasible result.

            \begin{algorithm}
                \caption{Communication Constraint Search}
                \label{alg_CCS}

                \KwIn{Map $\mathcal{M}$}
                \KwIn{Ordered Sequence of Goals $\mathcal{G} = \{ g_1 \ldots g_n\}$ }
                \KwIn{Number of agents $N$}
                \KwIn{Initial positions $p$}
                \Parameter{Communication Limit $\zeta$}

                \tcc{Stage 1}

                $R \leftarrow \{\emptyset_{1}, \ldots, \emptyset_{N}\}$; \\ \label{alg_stage1_a}
                $\mathcal{S} \leftarrow \{\}$; \\
                \For{$g \in \mathcal{G}$}{ \label{alg_foreach_epoch_s1}
                    $l, pos \leftarrow \text{SelectLeader}\left(R, g\right)$; \\ \label{alg_leader_select}
                    $\pi \leftarrow \text{SelectOrder}\left(pos, R\right)$; \\ \label{alg_follow_order}
                    $R\left[l\right] \leftarrow \text{InitializeAgent}\left(pos, \mathcal{M}\right)$; \\
                    $\text{PlanPath}\left(R\left[l\right], g\right)$; \\ \label{alg_leader_plan}
                    \For{$i \in 1\ldots \left(N - 1\right)$}{
                        $o \leftarrow \pi\left[i\right]$; \\
                        $cur \leftarrow \text{InitFromPrev}\left(R, o, \pi\left[0\ldots i - 1\right]\right)$; \\ \label{alg_init_from_prev}
                        $\text{FollowerPlan}\left(cur, R, \pi\left[0\ldots i - 1\right]\right)$; \\ \label{alg_follower_plan}
                        $R\left[o\right] \leftarrow cur$;
                    }
                    $\text{ValidateResult}\left(R\right)$; \\ \label{alg_validate}
                    $\mathcal{S} \leftarrow \mathcal{S} \cup \{R\}$; \\
                    \label{alg_stage1_b}
                }
                \hrule height 3pt
                \tcc{Stage 2} \label{alg_stage2_a}
                $initpos \leftarrow p$; \\
                $inittimes \leftarrow \{0\ldots 0\}$; \\
                $\mathcal{P} \leftarrow \{\}$; \\
                \For{$s \in \mathcal{S}$}{
                    $\mathcal{C} \leftarrow \text{InitCompositeSearch}\left(initpos, inittimes\right)$; \\
                    \For{$i \in 0\ldots \left(N - 1\right)$}{ \label{alg_compute_heuristic_a}
                        $b \leftarrow \text{PlanLinkedPath}\left(s\left[i\right].result, \emptyset, \emptyset\right)$; \\
                        $\mathcal{C}.heuristic\left[i\right] \leftarrow b.nodes$; \\ \label{alg_compute_heuristic_b}
                    }
                    $\mathcal{C}.goal \leftarrow \left(s.leader, s.goal\right)$; \\
                    $\text{RunCommunicationConstrainedBFS}\left(\mathcal{C}\right)$; \\ \label{alg_cc_bfs}
                    $initpos \leftarrow \mathcal{C}.result.position$; \\
                    $inittimes \leftarrow \mathcal{C}.result.costs$; \\
                    $\mathcal{P} \leftarrow \mathcal{P} \cup \{\mathcal{C}.result\_path\}$; \\
                }
                \label{alg_stage2_b}
            \end{algorithm}

            At the start of each epoch, for each agent, there is a set of nodes it could be at.
            In order to select the leader, the algorithm performs a single-source search from the vertex $g_i$ until it reaches all the result nodes of the previous epoch.
            This way, the search finds the agent that is (could be) closest to the current goal and fixes its location.
            Importantly, this also discards some result nodes in the previous epochs (those that could not support the selected leader starting vertex, this happens during initialization of the search on \Cref{alg_init_from_prev}).
            Since the leader $\agent_{l_i}$ does not take into account the communication constraint, it can plan its path right away (\Cref{alg_leader_plan}).

            Once the leader is selected, an ordering of the follower agents is established (\Cref{alg_follow_order}).
            This is done by repeatedly running a multi-source shortest paths computation.
            In the first iteration, only the leader starting location is added as a source node.
            The search then finds the closest agent that is not yet in the ordering.
            Once that agent is found, a new multi-source shortest path search is initialized.
            Its source nodes are set to all the source nodes from the previous search.
            Then, all the potential starting locations of the newly added agent are checked.
            Any that can communicate with the source nodes of other agents already in the new search are added as new source nodes for the next iteration.
            This way, the order respects the topology of the communication tree and an infeasible order cannot be selected.
            The followers are then arranged in the order in which their starting locations were added to the search.
            The notation $\pi_i$ is used to denote the order of the agents in epoch $i$.
            It always holds that $\agent_{\pi_i\left[0\right]} = \agent_{l_i}$, i.e., the leader is the first in the ordering.

            \begin{figure*}[t]
                \newcommand{\mwdth}{0.48}
                \centering
                \begin{subfigure}{\mwdth\linewidth}
                    \includegraphics[width=\linewidth]{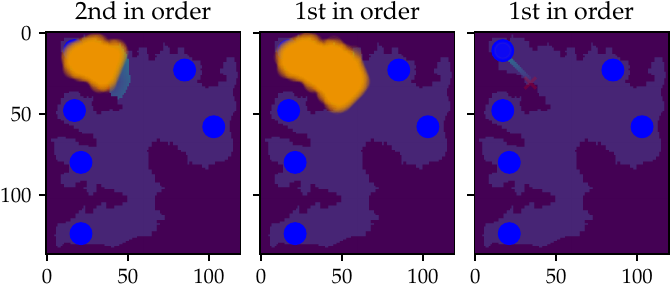}
                    \caption{Epoch 1}
                \end{subfigure}
                \begin{subfigure}{\mwdth\linewidth}
                    \includegraphics[width=\linewidth]{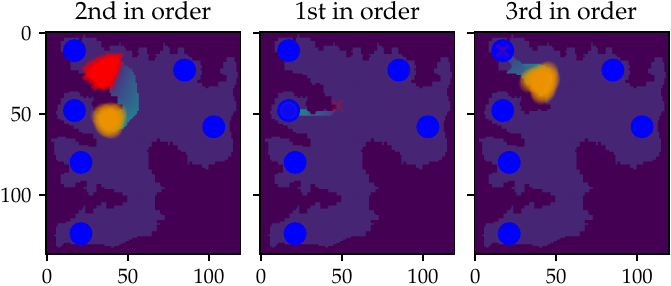}
                    \caption{Epoch 2}
                \end{subfigure}
                \begin{subfigure}{\mwdth\linewidth}
                    \includegraphics[width=\linewidth]{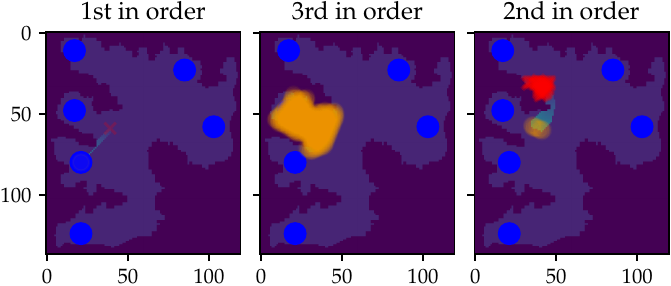}
                    \caption{Epoch 3}
                \end{subfigure}
                \begin{subfigure}{\mwdth\linewidth}
                    \includegraphics[width=\linewidth]{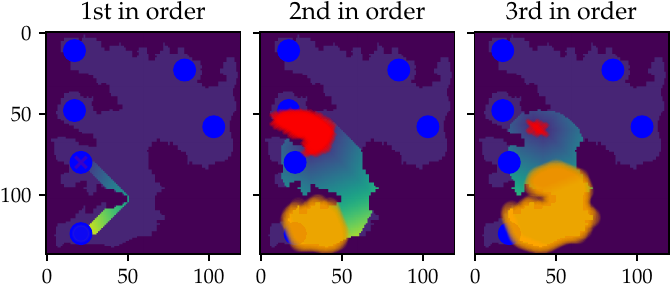}
                    \caption{Epoch 4}
                \end{subfigure}
                \begin{subfigure}{\mwdth\linewidth}
                    \includegraphics[width=\linewidth]{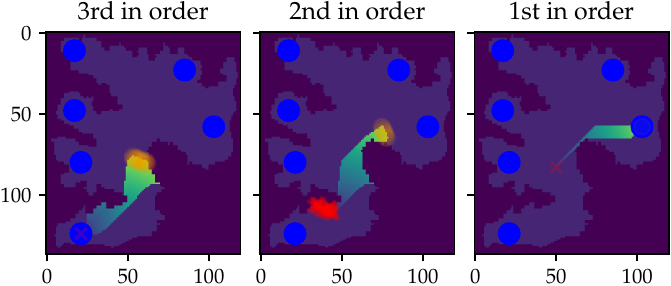}
                    \caption{Epoch 5}
                \end{subfigure}
                \begin{subfigure}{\mwdth\linewidth}
                    \includegraphics[width=\linewidth]{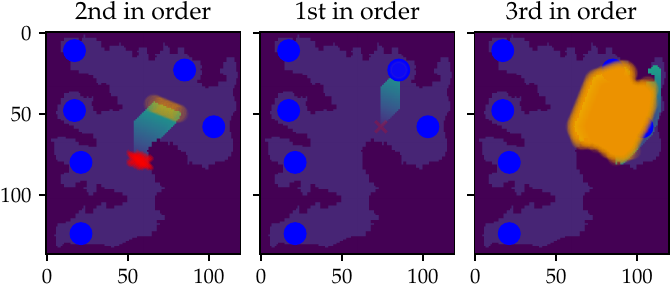}
                    \caption{Epoch 6}
                \end{subfigure}
                \caption{Example of Stage 1 computation for a simple map. Darkest color represents obstacles. Blue dots represent goals. The shaded areas show the nodes explored by the agent (lighter color corresponds to nodes opened later in time). Orange shading shows result nodes, red shading (crosses) represents potential starting nodes (sometimes not visible under the result node marks).}
                \label{fig_stage1}
            \end{figure*}

            The follower agents run their searches in this order, validating the communication constraint with respect to the already planned paths and ignoring the agents later in the ordering.
            The search is initialized with all the result nodes of the previous epoch that can communicate with the initial nodes of the previous agents.
            The search is then almost the same as a regular shortest-path search.
            The only difference is that each node has to also be checked for communication.
            For a node to be valid, there must already exist a \key{witness} node of an agent earlier in the ordering with roughly the same cost that the current node can communicate with (the cost cannot be matched exactly since we have non-uniform edge weights).
            The agent $\agent_{\pi_i\left[n\right]}$ validates only against $\agent_{\pi_i\left[0\right]}\ldots\agent_{\pi_i\left[n - 1\right]}$.
            If a witness exists, the current node is valid, regardless of which agent the witness belongs to.

            For each of the agents, some of the search nodes are considered to be \key{result nodes}.
            Those represent the nodes the agent can be at when the epoch ends (when the goal has been reached by the leader and all agents stopped their motions).
            For the leader, this is just the node that visits the goal.
            For the followers, any node that can reach a result node of a previous agent without breaking communication is also considered a result node.
            Notice that this is a stronger condition than just maintaining communication.

            If it happens that a follower search concludes without any result nodes, the follower agent requires the other agents to wait for it to catch up.
            This is done by decreasing the costs of the frontier search nodes (nodes without children).
            Each node that had its cost changed is reinserted into the queue and the search is started again.
            This is repeated until the search finishes successfully.
            Since the result nodes from the previous epoch are reachable from each other without breaking communication, the worst-case scenario is that the waiting step is repeated until the agent reaches the starting location of another agent and then follows the exact same path.

            After all of the follower agents finish their searches, there is a backward validation step (\Cref{alg_validate}).
            Since the communication has been checked only with respect to previous agents, some of the result nodes may be infeasible with respect to the agents later in the order.
            There may be result nodes for an early follower that cannot support any of the result nodes of the later nodes.
            Thus, if such a node was selected, the later agents could not keep the communication, and the algorithm would fail.
            
            To solve this, the result nodes are searched in the opposite order: all the result nodes of the last follower are automatically valid, and for every previous agent, the result nodes are only valid if they are (could be) a witness to at least one result node of a later agent.
            Still, this approach on its own does not cover the issue.
            Since the communication graph could be a tree, any leaf agents do not need to support any other agent to be valid, and a lot of them cannot support any of the later agents.
            Additionally, we do not know the exact structure of the graph, we only have a topological ordering of the agents.
            If, after the first pass, there are any agents that do not have any valid result nodes, we assume those to be on the leaves of the communication tree.
            To account for those, the backward search runs again from the last such agent, and continues again to the root.
            This step is repeated until all agents are connected.

            Even with this addition, this validation could still lead to a deadlock between the agents.
            If one agent supports two different agents, but none of its result nodes can support both of them at the same time, the validation step might still cause the algorithm to fail.
            Our solution does not handle this edge case, since it is highly unlikely that it actually occurs.

            At the end of an epoch, we have a set of possible locations for each agent.
            The next epoch then initializes its searches with these locations (\Cref{alg_init_from_prev}).
            Each of those nodes can be reached without breaking the communication (since it has a witness), but the algorithm does not determine how.

            \begin{figure*}[t]
                \centering
                \newcommand{\hgh}{5.6cm}
                \begin{subfigure}{0.33\linewidth}
                    \includegraphics[height=\hgh]{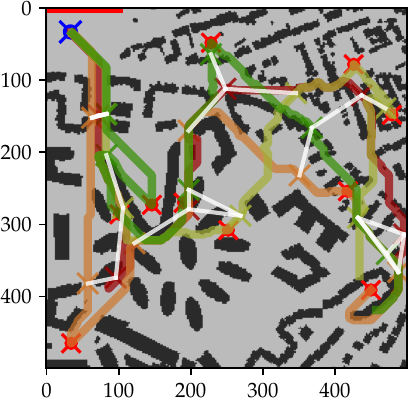}
                \end{subfigure}
                \begin{subfigure}{0.30\linewidth}
                    \includegraphics[height=\hgh]{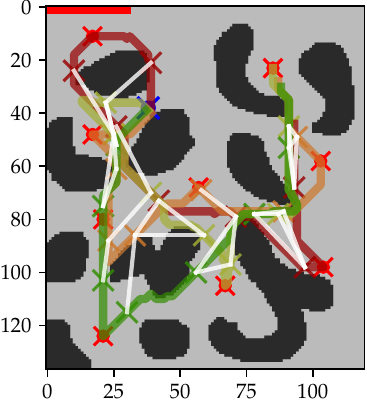}
                \end{subfigure}
                \begin{subfigure}{0.33\linewidth}
                    \includegraphics[height=\hgh]{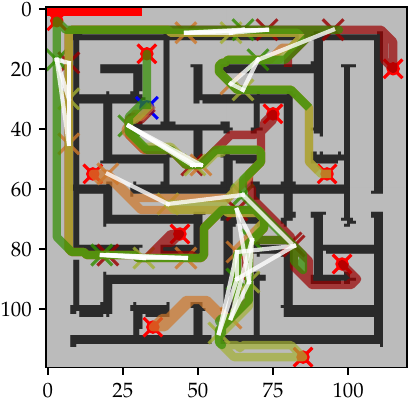}
                \end{subfigure}
                \caption{Maps: Paris, (left to right) Blobs, Maze. Red and blue marks show goals and starting position (agents start close together around this point). Traces of the agents shown in colors. Potential communication links in white, communication limit in red in the top left corner. Each white cluster represents a different time snapshot, and represents the full communication graph.}
                \label{fig_maps}
            \end{figure*}

            \Cref{fig_stage1} shows the state of computation in Stage 1 at the end of each epoch on an example of a simple map.
            The agents later in the communication chain typically have a wider area they can cover, which puts them in more favorable positions with respect to the next goal.
            Notice how the potential starting locations (shown in red) are only a small subset of the result nodes from the previous epoch (shown in orange).
            This is because the selection of the leader fixes the starting location of that agent in one location and all other agent must respect that.
            The selected leader had most likely more than one result node.
            Therefore, all the result nodes that cannot be supported with the selected leader position must be discarded.

            Notice also the second epoch.
            Because the leader was able to maneuver close to the goal during the previous epoch, the epoch was very short.
            Even so, the other agents were able to improve their location with respect to the next goal.
            On more complicated maps, it is not uncommon for an epoch to have zero length, since the leader already maneuvered to the target goal in the previous epoch.
            This highlights, how the algorithm keeps all the potential options open as long as possible, which allows it to take into account goals in future epochs while keeping the search focused to a single goal.

        \subsection{Stage 2}

            Once the first stage finishes, each agent has a search tree for each epoch, and all the nodes in those trees are reachable.
            Each agent also has a set of potential result nodes that are advantageous with respect to the future epochs
            Because the leader selection discards useless nodes and the follower agents need to follow the leader, any of the result nodes left at the end of stage 1 should provide a good starting point for the next epoch.

            The second stage uses these results to compute a greedy best-first-search.
            The heuristic function used can then be defined as follows:
            \begin{equation}
                w\left(a_0,\ldots,a_{n - 1}\right) = \sum_{i = 0}^{n - 1}\left(c_i\left(a_i\right)^\alpha\right),
                \label{eq_heuristic}
            \end{equation}
            where $c_i\left(a_i\right)$ is a simple cost-to-reach for the agent $a_i$ to the closest of its result nodes, and $\alpha > 0$ is a manually set parameter (value of \num{1.5} yielded best results in our experiments).
            The heuristic computed by \Cref{eq_heuristic} can, in some cases, yield same results for two actions.
            In such cases, the search is biased towards the goal.
            The computation of the heuristic is shown in \Cref{alg_CCS}, \Crefrange{alg_compute_heuristic_a}{alg_compute_heuristic_b}.

            The cost-to-reach is computed using a multi-source shortest-path search from the result nodes.
            Additionally, the search is confined to the vertices opened by the search in stage 1.
            Any other vertices have been either too far or in violation of the communication constraint. 
            Those nodes are assigned a very high heuristic value.

            The search itself is then a best-first-search based on the heuristic computed in this manner, performed in the composite space of the agents (\Cref{alg_CCS}, \Cref{alg_cc_bfs}).
            It is initialized with the starting times and starting locations of each agent.
            When a node is expanded, the agent with the lowest current time takes action - either a move or a waiting action.
            When an agent moves, its location changes and its time is increased by the weight of the taken edge.
            When it waits, its time is increased to match the time of the first non-waiting agent (which corresponds to the first time instance when the circumstances change), and so is that of all other waiting agents.

            If an agent moves in a way which breaks the communication links with agents that could still move, the action is taken, the node is marked as speculative, and the time of the new action is noted.
            All of its children nodes are going to be marked as speculative with, and the time stamp cannot decrease.
            A node stops being speculative, if the communication constraint is met.
            A node is only discarded, when the node is in violation of the constraint and the current time of the last agent is at or above the speculation time stamp.
            Essentially, the solver can accept an action \textit{conditioned} by the actions of the other agents.
            This way, actions that would require two agents to act simultaneously can be performed, even though the actions in the search are explored sequentially and would be infeasible individually.

            The composite-state search proceeds a single epoch at a time, and when the current goal in the epoch is reached, the next epoch is initiated immediately with the result position and time as the initial search node.
            Notice that only the leader reaching the goal is required for the epoch to finish.
            It does not matter where any of the other agents are as long as they are within communication distance, even if that agent is the leader for the next epoch.
            The search continues until all the goals have been visited.
            The result of the algorithm is then a sequence of actions for each of the agents, such that at any point in the plan, any agent is able to communicate (in terms of the relation $\hat{\mathcal{C}}$ defined in \Cref{sec_problem}) with any other agent.

    \section{Experiments and Results}
        \label{sec_results}

        \begin{figure}[t]
            \centering
            \includegraphics[width = \linewidth]{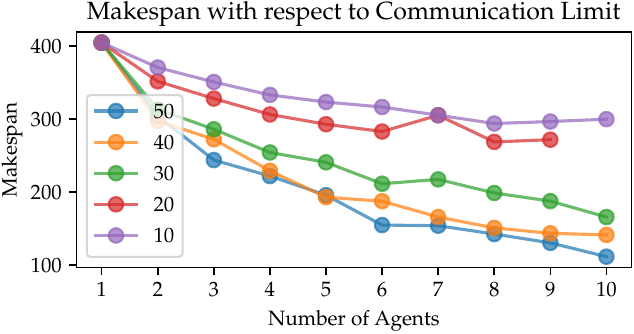}
            \caption{The makespan with respect to the number of agents and the communication range for the map Blobs (column with 1 agent is just a single agent shortest tour visiting all the goals for comparison). Missing value is the instance where none of the experiments finished successfully.}
            \label{fig_costs}
        \end{figure}
        
        \begin{figure}[t]
            \includegraphics[width=\linewidth]{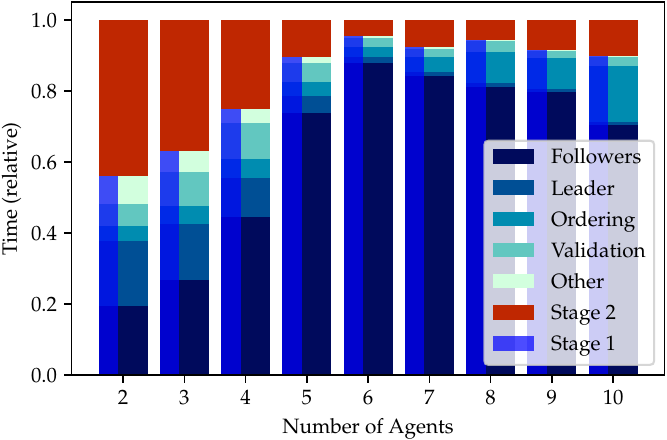}
            \caption{Time requirements of the individual computational steps, normalized. Showing result on the Blobs map, average values over all tested communication limits.}
            \label{fig_time_portions}
        \end{figure}

        \begin{figure}[t]
            \includegraphics[width=\linewidth]{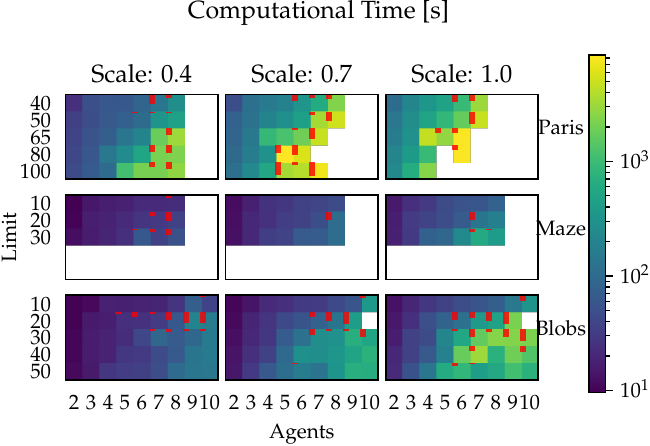}
            \centering
            \caption{Time requirements with respect to the number of agents and the communication limits. Time in seconds, logarithmic scale, results for scale 0.4, 0.7 and 1 (Note that 0.4 scale Paris is still much larger than full scale Blobs, and 0.4 scale Maze does not look like a maze at all (the map just cannot preserve its structure when scaled)). Red bar represents percentage of failed computations.}
            \label{fig_time_grid}
        \end{figure}

        We tested our algorithm on a number of different maps from~\cite{pathplanning_data} (adjusted in scale) and some manually created ones.
        Some of the used maps are shown in \Cref{fig_maps}.
        This section discusses the computational intensity of the algorithm with respect to different aspects - graph size, number of agents and communication limit.

        We assumed an 8-neighborhood with edge costs of $1$ and $1.4$, and the communication distance function $\gamma$ was defined as the Euclidean distance.
        The goals were placed into the map manually, and the ordering was obtained using a plain TSP solver.
        In order to test the computational properties with respect to graph sizes, we ran the experiments on scaled instances.
        Additionally, instead of having fixed starting locations, the initial positions were placed randomly around the specified location.
        This randomness allowed us to run the same experiment several times with very slightly different parameters, giving us more data about the success rate of the algorithm.
        Each experiment ran multiple times (a total of over \num{5000} instances) and had a timeout of \SI{10000}{\second} on a machine with Intel Xeon Gold 6130 CPU \SI{2.10}{\giga\hertz}, single core.

        \Cref{fig_costs} shows the results in terms of the makespan, which was the minimized criterion and can be seen as a metric of the quality of the proposed solution.
        As the figure shows, overall, the length of the resulting plan goes significantly down, both when the number of agents increases as with increasing communication range.
        The figure shows that adding more agents has diminishing returns.
        This makes sense - increasing the number of agents gives more options to the next leader.
        Consequently, the duration of the epoch decreases, which partially counteracts this advantage.
        Also, the hard lower bound on the makespan is the distance from the initial locations to the goals, no matter how many agents are used, the result cannot be lower than this.

        Notice also that there are cases where increasing the number of agents resulted in an increase in the resulting makespan.
        The Stage 2 of the algorithm can sometimes struggle with local minima in the map, which can cause it to produce worse results when the constraints are loosened.

        In terms of the computational demands, the proposed algorithm is quite intensive.
        Most of the presented experiments were done using the maps named Blobs (120 x 140, up to 10 agents), Paris (500 x 500, up to 8 agents) and Maze (120 x 120, up to 8 agents, fewer tested communication limits on this map, since the maze structure does not really lend itself to cooperation between the robots and the result are not as interesting; still, it highlights the capabilities of the solver).
        As is shown in \Cref{fig_time_grid}, the algorithm is capable of handling large communication limits (relative to the size of the map) and can handle up to 10 agents, although the computational times are quite significant, with many of the instances requiring an hour or more to compute.
        At the same time, the number of nodes is one of the most important factors, and the algorithm works on general graphs.
        Therefore, it can be used, for example, with roadmaps, which typically do not have as high number of nodes (non-scaled version of the map called Paris is equivalent to \num{250000} nodes).

        \Cref{fig_time_portions} shows how much of the computation the individual steps of the algorithm take.
        The vast majority of the computational complexity comes from the first stage, and the ratio increases even more as the number of agents increases.
        This shows how effective the heuristic is - if the computation is successful, the complexity of composite state space in the second stage is close to negligible in the overall computation.
        Within the first stage, most time is spent in the follower agent search part of the algorithm, which corresponds to \Cref{alg_follower_plan} in \Cref{alg_CCS}.
        This boils down to the communication constraint check within the search.
        Therefore, even though the algorithm should work with any distance metric, the computational burden of computing this metric could be prohibitive (which, on the other hand, could be overcome by precomputing all-pairs distances between all the nodes).

        Interestingly, with higher agent counts, the portion of time spent ordering the agents at the start of each epoch increases significantly.
        Again, this is to some extent caused by the communication limit, which necessitates additional checks to make sure that the selected ordering is actually feasible.

    \section{Conclusion}
        \label{sec_conclusion}

        This work presents the Communication-Constrained Multi-Agent Path-Planning problem and proposes a novel centralized algorithm based on graph search to address this problem.
        The problem consists of finding a path in an environment represented by a graph, such that multiple goals are visited, and the makespan of the entire mission is minimized.
        The proposed approach is capable of handling swarms of up to ten agents, potentially more, depending on the size of the map and the communication range.

        The algorithm was tested on several 8-neighborhood grid-world scenarios, and showed potential for implementation in tandem with a roadmap-based algorithm.
        The approach is generic and should be able to incorporate different communication range metrics, which we hope to focus our future research on.

    \section*{Acknowledgment} 

        This research was funded by Czech Science Foundation grant number 22-30043S.

    \bibliography{main}
    \bibliographystyle{unsrt}

\end{document}